\UseRawInputEncoding
\documentclass[journal]{IEEEtran}
\usepackage{soul}

\usepackage{cite}

\usepackage[pdftex]{graphicx}
\graphicspath{{../Drawings/}{./Drawings/}{./}}
\DeclareGraphicsExtensions{.pdf}

\usepackage[usenames, dvipsnames]{color}

\usepackage{amsmath,amssymb}
\usepackage{bbm}
\renewcommand\Re{\operatorname{Re}}
\renewcommand\Im{\operatorname{Im}}

\usepackage[caption=false,font=footnotesize]{subfig}

\usepackage{url}

\begin{document}

\title{Nonuniform Sampling Rate Conversion: \\An Efficient Approach}

\author{Pablo~Mart\'inez-Nuevo,~\IEEEmembership{Member,~IEEE}
\thanks{\textcopyright\ 2021 IEEE. Personal use of this material is permitted. Permission from IEEE must be obtained for all other uses, in any current or future media, including reprinting/republishing this material for advertising or promotional purposes, creating new collective works, for resale or redistribution to servers or lists, or reuse of any copyrighted component of this work in other works.}
\thanks{P. Mart\'inez-Nuevo is with the research department at Bang \&~Olufsen, 7600 Struer, Denmark (e-mail: pmnuevo@alum.mit.edu).}% <-this % stops a space
\thanks{Digital Object Identifier 10.1109/TSP.2021.3079802}}

\markboth{IEEE Transactions on Signal Processing}%
{Shell \MakeLowercase{\textit{et al.}}: Bare Demo of IEEEtran.cls for IEEE Journals}

\maketitle

% 150-250 words for transactions on signal processing

\begin{abstract}
We present a discrete-time algorithm for nonuniform sampling rate conversion that presents low computational complexity and memory requirements. It generalizes arbitrary sampling rate conversion by accommodating time-varying conversion ratios, i.e., it can efficiently adapt to instantaneous changes of the input and output sampling rates. This approach is based on appropriately factorizing the time-varying discrete-time filter used for the conversion. Common filters that satisfy this factorization property are those where the underlying continuous-time filter consists of linear combinations of exponentials, e.g., those described by linear constant-coefficient differential equations. This factorization separates the computation into two parts: one consisting of a factor solely depending on the output sampling instants and the other consists of a summation---that can be computed recursively---whose terms depend solely on the input sampling instants and its number of terms is given by a relationship between input and output sampling instants. Thus, nonuniform sampling rates can be accommodated by updating the factors involved and adjusting the number of terms added. When the impulse response consists of exponentials, computing the factors can be done recursively in an efficient manner.
\end{abstract}


% \fi
\IEEEpeerreviewmaketitle

\section{Introduction}
\IEEEPARstart{T}{he} interconnection of different digital discrete-time systems operating at different rates makes sampling rate conversion a fundamental operation in most modern signal processing chains. This operation can be seen as resampling after reconstruction and, in principle, it is possible to perform it in the continuous- or discrete-time domains \cite{Oppenheim:2010aa,Proakis:2009aa}. However, in many practical applications, it is convenient that this conversion is carried out entirely in the discrete-time domain. One example is within the context of audio signal processing where most of the processing and interfacing between digital data streams is performed in discrete time. Thus, on the one hand, it makes the interconnection simpler and more flexible and, on the other hand, avoids the distortion caused by the D/A and A/D converters. 

From a practical point of view, sampling rate conversion in the discrete-time domain has been further classified into synchronous and asynchronous \cite{Adams:1994aa,Beckmann:2005aa}. These definitions are inspired by the clock configuration setting the rates of the system. In synchronous sampling rate conversion, it is assumed that there exists a single master clock where the different rates of the system are related to by a fixed factor. This factor is generally considered to be a rational number. The asynchronous counterpart assumes several separate clocks operating at different rates. This situation gives rise to sampling rates related by an arbitrary factor that can also change over time. 

From a conceptual point of view, sampling rate conversion techniques commonly tackle the problem from two sides. First, if the sampling rates are related by a constant rational factor, there exist efficient algorithms that take advantage of this relationship within the context of filter banks \cite{Crochiere:1983aa,Vaidyanathan:1993aa}. The conversion in this case consists of upsampling followed by downsampling both by integer factors \cite{Oppenheim:2010aa}. When both of these factors become large, the computational requirements become very demanding. This situation often arises when the values of input and output sampling rates are relatively close, e.g., from 44.1~kHz to 48~kHz or for two systems with the same nominal rate presenting a small deviation between them. 

In the second approach, the sampling rates are considered arbitrary, i.e., they are not assumed to have a particular relationship between them and are allowed to vary with time. Note that this potential fluctuation of the conversion ratio implies that both the input and output samples can correspond to nonuniform samples of the underlying continuous-time signal. We refer to this as nonuniform sampling rate conversion.

There are many applications that can generate nonuniform samples. For example, multichannel data acquisition \cite{Zhao:2015aa}, data loss in communication networks \cite{Bakri:2018aa}, synchronization errors in interleaved A/D converters \cite{Nikaeen:2009aa}, and deliberate nonuniform sampling for data compression \cite{Mark:1981aa}. Nonuniform sampling rate conversion becomes particularly relevant whenever nonuniformly or uniformly sampled data, typically coming from different sources, have to lock to a common master clock, e.g., digital audio \cite{Adams:1993aa}, satellite communications \cite{Takahata:1987aa}, or synchronous networks \cite{Sinha:2016aa}. It can additionally be found in particle accelerators \cite{Guarch:2020aa}.

Arbitrary sampling rate conversion entirely in the digital domain is commonly addressed by means of a time-varying discrete-time filter. This approach can be shown to be equivalent to resampling after reconstruction, i.e., reconstructing the sequence of samples as a continuous-time signal and then sampling again at a different rate. The challenge lies in efficiently updating the filter coefficients at each time step and performing the corresponding filtering operation. A common strategy in the literature is based on either storing a large number of samples and performing simple interpolation---e.g. first-order or cubic interpolation---or storing fewer samples at the expense of more sophisticated interpolation techniques \cite{Ramstad:1984aa,Smith:1984aa,Lagadec:1981aa,Lagadec:1982aa,Beckmann:2005aa}. In this approach, there is a tradeoff between computational complexity and memory requirements. 

In \cite{Russell:2002aa}, it is shown, for the case of proper rational transfer functions, a recursive computation of coefficients leading to low computational complexity and memory requirements. However, it still focuses on constant uniform input and output sampling rates. In \cite{Blok:2012aa,Blok:2014aa}, a technique is presented that can adapt to continuous changes in the conversion ratio. However, the tradeoff between computational complexity and memory requirements still stands similar to the approaches already mentioned. These algorithmic limitations still persist in specific hardware implementations. However, they can be somewhat alleviated, up to a certain degree, when exploiting a specific hardware design \cite{Adams:1994aa}. In particular, Farrow-based structures are becoming common in asynchronous or arbitrary sampling rate conversion \cite{Guarch:2020aa, Lijun:2014aa}.

In the literature detailed above, the approaches presented are, in general, focused on input and output sequences coming from a uniform sampling process---in some cases, allowing for slow changes in the conversion ratio. However, in this paper, we present a technique that accommodates the processing of sequences at nonuniform input and output rates which is also computationally efficient, i.e., low complexity and memory requirements. In other words, it can adapt to instantaneous changes of the sampling conversion ratio. This approach is based on the factorization properties of the underlying continuous-time filter. This is particularly advantageous when this filter consists of linear combinations of exponentials which is precisely the case of proper rational transfer functions. We show how the filtering can be arranged so that the computation of coefficients can be done recursively.

In the next section, we introduce nonuniform sampling rate conversion as filtering of nonuniform time samples. We show how this framework generalizes arbitrary sampling rate conversion. For conceptual purposes, we view this process as resampling after reconstruction which can be equivalently achieved in discrete-time by a time-varying filter. In Section \ref{section:NonuniformFiltering}, we show that some causal impulse responses, which can be factorized in a particular manner, may be amenable to efficient computation. In particular, we focus on impulse responses consisting of linear combinations of exponentials, e.g., those having rational transfer functions. Section \ref{section:FirstOrder} analyzes the algorithm for first-order systems and Section \ref{section:NOrderSystems} extends it to repeated poles on the real axis. In Section \ref{section:SecondOrderSystems} we consider systems with distinct poles in complex conjugate pairs. We conclude the paper by illustrating the algorithm with an example.

For ease of notation, we refer to $x[n]$, $n\in\mathbb{Z}$, as discrete-time signals or sequences of samples whose values can arrive at uniform or nonuniform instants of time. If samples arrive at arbitrary instants of time, we refer to the corresponding signals as nonuniform sequences and we assume that they may have originated from a nonuniform sampling process. Similarly, we use the term uniform sequences if the sample values arrive at equally-spaced instants of time. We also assume that these may have originated from a uniform sampling process.

%\IEEEpubidadjcol

\section{Nonuniform Sampling Rate Conversion}
\label{section:NonuniformSRC}
\begin{figure}[thpb]
\centering
\includegraphics[width=.85\columnwidth]{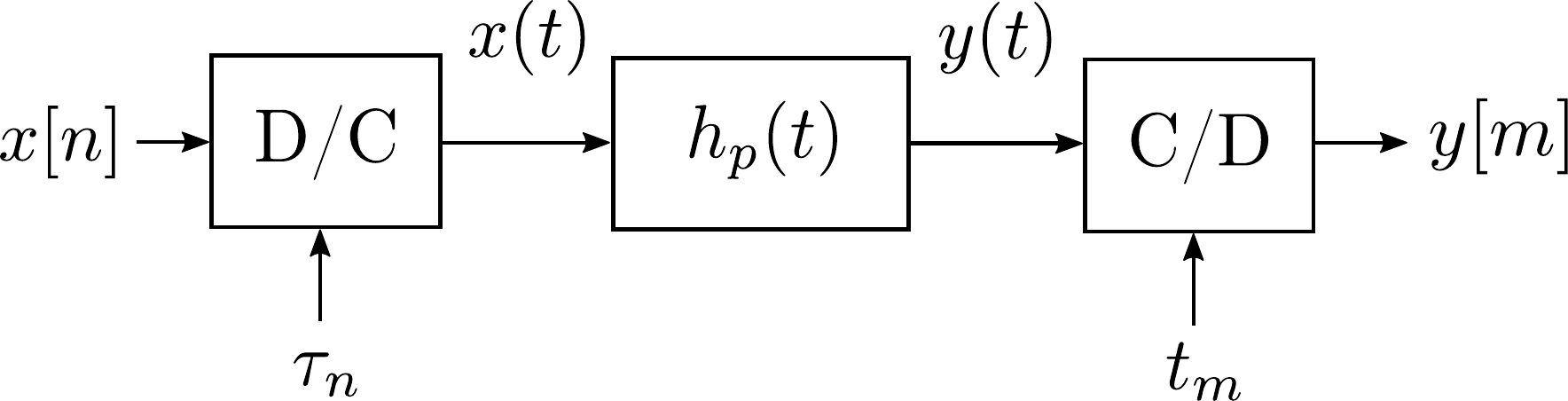}
\caption{Nonuniform sampling rate conversion viewed as reconstruction and resampling in the continuous-time domain.}
\label{fig:NonuniformSRC}
\end{figure}

Sampling rate conversion can be better understood by considering the reconstruction of the discrete-time signals involved in the process. In particular, we can associate a continuous-time signal $x(t)$ to a given discrete-time sequence by assuming that $x(t)$ is constructed from a sequence of samples. In particular, consider $x[n]:=x(\tau_n)$ where $\tau_n:=nT_x\epsilon_{x,n}$ for some $T_x>0$, $\epsilon_{x,n}\in\mathbb{R}$, and $n\in\mathbb{Z}$. If $\tau_0\neq0$ is required, it can be defined differently; however, we will maintain this definition for notational convenience. We can then write
\begin{equation}
\label{eq:NonUniSRC}
x(t)=\sum_{n\in\mathbb{Z}}x[n]h_r(t-\tau_n).
\end{equation}
for some finite-energy $h_r$. Typically, classical nonuniform sampling reconstruction series may not take the form of a linear combination of time shifts of $h_r$ \cite{Levinson:1936aa}. Thus, in practice, the reconstruction in (\ref{eq:NonUniSRC}) can be approximated by a sinc-like interpolation \cite{Maymon:2011aa}. If we now want to resample this continuous-time signal $x(t)$ at time instants $t_m:=mT_y\epsilon_{y,m}$ for some $T_y>0$, $\epsilon_{y,m}\in\mathbb{R}$, and $m\in\mathbb{Z}$, it may be desirable to perform some processing before---e.g. in order to avoid aliasing---and sample the signal $y(t):=(x*h_p)(t)$ instead. Then, nonuniform sampling rate conversion consists of taking $x[n]$ and generating $y[m]:=y(t_m)$ for $n,m\in\mathbb{Z}$. In principle, this process can be carried out in the continuous-time domain by considering it as resampling after reconstruction of $x(t)$ from the nonuniform samples $x[n]$. Note that if $h_p(t)$ is an ideal lowpass filter whose cutoff frequency is at the Nyquist frequency with respect to $x(t)$, the system outputs the nonuniform samples $\{x(t_m)\}_{m\in\mathbb{Z}}$. This process is illustrated in Fig.~\ref{fig:NonuniformSRC}. We assume that both the input and output sampling instants form a strictly increasing sequence. This is guaranteed if
\begin{equation} 
\epsilon_{\cdot,l+1}>\epsilon_{\cdot,l}\frac{l}{l+1}
\end{equation}
for all $l\in\mathbb{Z}$.

\begin{figure}[thpb]
\centering
\includegraphics[width=.7\columnwidth]{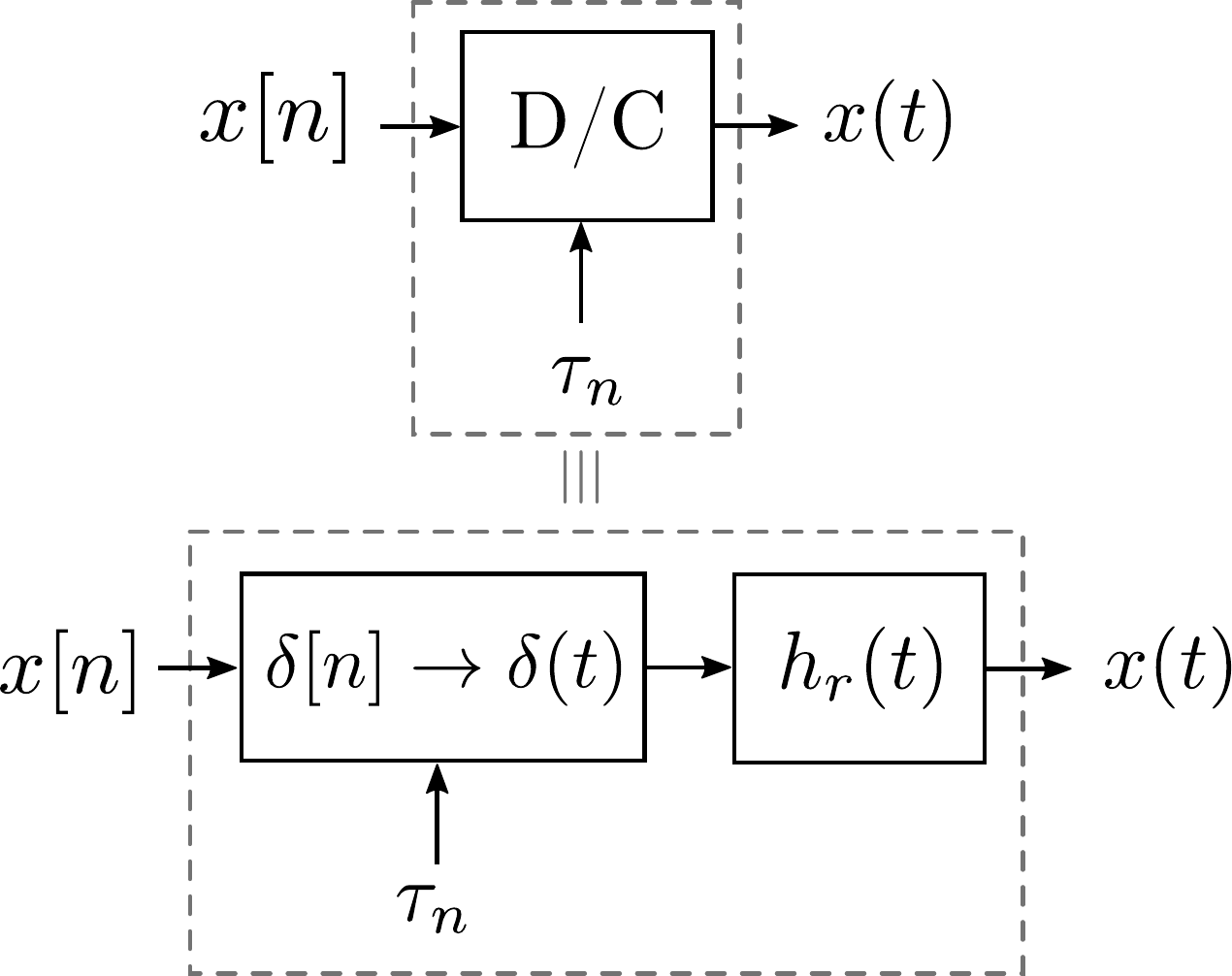}
\caption{Representation of the discrete-to-continuous block as a sequence-to-impulses operation followed by a reconstruction filter.}
\label{fig:DCblockNonUni}
\end{figure}

The continuous-to-discrete block, denoted by C/D, takes a continuous-time signal $y(t)$ and outputs its samples $\{y(t_m)\}_{m\in\mathbb{Z}}$. The operation of the discrete-to-continuous (D/C) block can be further split into two parts. Fig.~\ref{fig:DCblockNonUni} shows its two components. The first block converts the sequence to an impulse train, i.e., it outputs the continuous-time signal $\sum x[n]\delta(t-\tau_n)$. The resulting signal is then passed through a reconstruction filter $h_r(t)$. We could combine the reconstruction filter and the filter for additional processing $h_p(\cdot)$ into a single filter $h(t):=(h_r*h_p)(t)$. Then, the signal that is resampled after reconstruction, i.e., $y(t)=\sum x[n]h(t-\tau_n)$, generates the samples
\begin{equation}
\label{eq:SamplesOut}
y[m]=\sum_{n\in\mathbb{Z}}x[n]h(t_m-\tau_n).
\end{equation}

\begin{figure}[thpb]
\centering
\includegraphics[width=.6\columnwidth]{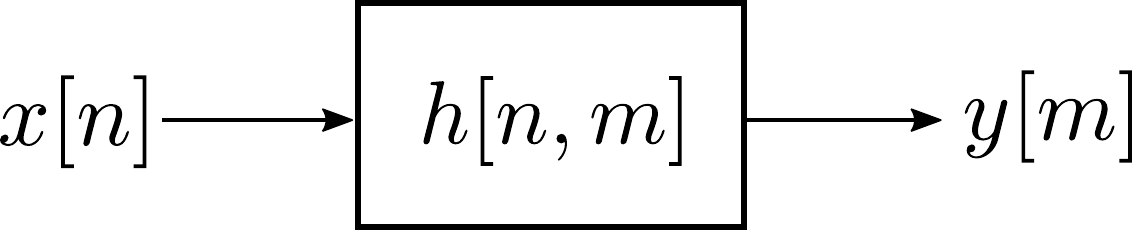}
\caption{Time-varying discrete-time system for nonuniform sampling rate conversion.}
\label{fig:DT_NonUniSRC}
\end{figure}

From (\ref{eq:SamplesOut}), it can be seen that the entire conversion process can be carried out in the discrete-time domain by considering $h[n,m]:=h(t_m-\tau_n)$ as a time-variant discrete-time system (see Fig.~\ref{fig:DT_NonUniSRC}). Since the sampling process can be nonuniform, sampling sequences and cutoff frequencies can be framed under the concept of the Landau rate \cite{Landau:1967aa}.

Note that the coefficients $h[n,m]$ correspond to nonuniform samples of $h(\cdot)$. In the next sections, we show how the computation of these samples can be done efficiently with low memory and computational requirements when $h(\cdot)$ consists of a linear combination of exponentials.

\subsection{Uniform Sampling Rate Conversion}
Uniform sampling rate conversion can be seen as a particular case of the nonuniform case, i.e., $\epsilon_{x,n}=\epsilon_{y,m}=1$ for all $n,m\in\mathbb{Z}$. In this setting, the reconstruction filter can be considered an ideal filter performing bandlimited interpolation given by $h_r(t)=\mathrm{sinc}(t/T_x)$. Then, $h_p(t)$ can be taken as an antialiasing filter appropriately chosen according to the new sampling rate. Therefore, it is common in practice to choose the filter $h(t)=(h_r*h_p)(t)$ as a causal filter approximating an ideal one with cutoff frequency $\min\{\pi/T_x,\pi/T_y\}$ rad/s. The uniform samples of the signal $y(t)$ can then be written as
\begin{equation}
\label{eq:UniformSRC}
y[m]=\sum_{n\in\mathbb{Z}}x[n]h(mT_y-nT_x)
\end{equation}
for $m\in\mathbb{Z}$ where $x[n]$ is the uniform input sequence. The conversion process can also be carried out in the continuous-time domain. Equivalently, it can be performed in the discrete-time domain by considering the time-varying filter $h[n,m]=h(mT_y-nT_x)$ \cite{Liu:1969aa,Crochiere:1981aa,Russell:2002aa}. The discrete-time approach is, in general, preferable since nonidealities such as the distortion introduced by the D/A converter and the quantization effects of the A/D converter are significant drawbacks \cite[Chapter 11]{Proakis:2009aa}. One of the advantages of the resampling after reconstruction approach in the continuous-time domain is that it does not make any assumption whatsoever of the relationship between the two sampling rates, i.e. the framework is valid for any two arbitrary sampling rates. 

Similarly to the nonuniform case, each output sample requires knowledge of the samples of the impulse response $\{h(mT_y-nT_x)\}_{n\in\mathbb{Z}}$. This set of samples is, in principle, different for each output sample and for changes in $T_x$ or $T_y$ when considering nonuniform rates. This requirement for generating a new set of samples for each output sample emphasizes why many earlier approaches have traded off memory for computation and vice versa. In this paper, we show how our approach can compute efficiently, in a recursive manner, and with low memory requirements, this new set of samples. This is valid for any input and output sample rate and for instantaneous changes in the sampling rates.

\section{Filtering of Nonuniform Time Samples}
\label{section:NonuniformFiltering}
The key aspect exploited throughout this paper for efficient nonuniform sampling rate conversion is based on a useful decomposition of the impulse response $h(\cdot)$. In particular, our approach is focused on filters that can be separated as follows
\begin{equation}
\label{eq:SeparationProperty}
h(t-\tau)=h_1(t)h_2(\tau)u(t-\tau)
\end{equation}
where $u(\cdot)$ is defined as $u(t)=0$ for $t<0$ and $u(t)=1$ for $t\geq0$. Note that rational discrete-time filters, commonly used in practice, satisfy this separation property, e.g., lowpass Butterworth, Chebyshev, or elliptic filters \cite[Chapter 7]{Oppenheim:2010aa}. This separation property, satisfied by the impulse response, allows us to express the output signal as
\begin{equation}
\label{eq:SeparationStep1}
y[m]=h_1(t_m)\sum_{n\geq0}x[n]h_2(\tau_n)u(t_m-\tau_n).
\end{equation}
where we assume that the input sample values are zero for negative time indexes, i.e., $x[n]=0$ for $n<0$. Note that the summation is finite since $u(t_m-\tau_n)\neq0$ only for $t_m\geq\tau_n$. Equivalently, let us denote
\begin{equation}
\lambda_{m,n}:=\Big\lfloor m\frac{T_y\epsilon_{y,m}}{T_x\epsilon_{x,n}}\Big\rfloor
\end{equation}
where $\lfloor \cdot\rfloor$ denotes the floor function. Thus, the summation consists of the values of $n$ such that $0\leq n\leq\lambda_{m,n}$. It is clear that there always exists some integer value $\lambda_m$ such that
\begin{equation}
\{n\in\mathbb{Z}:0\leq n\leq\lambda_{m,n}\}=\{n\in\mathbb{Z}:n=0,\ldots,\lambda_m\}.
\end{equation}
Then, we can now write
\begin{equation}
\label{eq:NonuniformSRC_Separation}
y[m]=h_1(t_m)\sum_{n=0}^{\lambda_{m}}x[n]h_2(\tau_n).
\end{equation}
The complexity of this expression reduces to computing the values at the corresponding nonuniform instants of both factors of $h(\cdot)$. This factorization allows us to separate the output sampling instants into the factor $h_1(t_m)$ and a second factor consisting of a summation. The terms of the latter depend solely on the input sampling instants and the number of these terms is given by a relationship between input and output sampling instants. We will show later how this calculation can be done recursively with the added complexity of an exponentiation whenever the corresponding $\epsilon_{\cdot,\cdot}$ are not constant.

\subsection{Continuous-Time Filters Described by Differential Equations}
\label{section:SystemsOfInterest}
An example of a class of filters that we will be focusing on throughout this paper are those described by linear-constant coefficient differential equations. The particular structure of the resulting impulse response, which can take the form of a linear combination of exponentials, is amenable to recursively computing the filter coefficients and adapting to instantaneous changes in the input and output sampling rates. The general form for an $N$-th order equation is given by
\begin{equation}
\label{eq:LTIdifferential}
\sum_{k=0}^{N}a_ky^{(k)}(t)=\sum_{l=0}^{M}b_lx^{(l)}(t).
\end{equation}
Under certain assumptions, this equation can be interpreted as describing a system with input $x(t)$ and output $y(t)$. In particular, assuming initial rest---i.e. if $x(t)=0$ for $t\leq t_o$, then $y(t)=0$ for $t\leq t_o$---the input-output relationship corresponds to a causal LTI system \cite[Chapter 3]{Oppenheim:1997aa}. Continuous-time linear filters are usually described in this manner and analyzed in the Laplace transform domain. In particular, they yield rational transfer functions of the form
\begin{equation}
H(s)=\frac{Y(s)}{X(s)}=A\frac{\prod_{k=1}^M(s-z_k)}{\prod_{k=1}^N(s-p_k)}.
\end{equation}
for $A, z_k, p_k \in\mathbb{C}$.

If we assume that poles are distinct and that $N>M$, it can be shown, by partial fraction expansion, that the inverse Laplace transform corresponds to an impulse response of the form
\begin{equation}
\label{eq:ParallelStructure}
h(t)=\sum_{k=1}^{N}a_ke^{\alpha_kt}u(t)=\sum_{k=1}^{N}h_k(t)
\end{equation}
where $a_k, \alpha_k\in\mathbb{C}$. Due to the linearity property of the convolution, the output of such a system is the sum of the outputs for each of the systems $h_k(\cdot)$. This allows us to perform the convolution in a parallel manner. We will also show the case of repeated real poles, which can serve as the basis to extend it to repeated complex poles. However, for ease of explanation, we will mainly focus on distinct poles.

\subsection{Coefficient Computation}
In order to illustrate how the computation is arranged, let us take as an example one of the terms in (\ref{eq:ParallelStructure}), i.e., $h_k(t)=a_ke^{\alpha_kt}u(t)$. The output of the sample rate converter corresponding to this signal path can be expressed as
\begin{equation}
\label{eq:FOexample}
y^{(k)}[m]=a_k(e^{\alpha_kT_y\epsilon_{y,m}})^m\sum_{n=0}^{\lambda_{m}}x[n](e^{-\alpha_kT_x\epsilon_{x,n}})^n.
\end{equation}
If we firstly assume that $\epsilon_{x,n}=\epsilon_{y,m}=1$, it is straightforward to see how the factor $(e^{\alpha_kT_y})^m=(e^{\alpha_kT_y})^{m-1}e^{\alpha_kT_y}$ can be recursively computed for each output sample from the value in the previous time step. The same applies to the coefficients $(e^{\alpha_kT_x})^n$. However, one of the key benefits of designing a sampling rate converter in this way is the ability to accommodate nonuniform input and output sampling instants---equivalently, to adapt to instantaneous changes in the sampling rates---with low computational requirements. These changes require computation of the factors $(e^{\alpha_kT_y\epsilon_{y,m}})^m$ and $(e^{-\alpha_kT_x\epsilon_{x,n}})^n$ as well as updating the number of terms in the summation.

The update of coefficients in both cases is carried out in the same manner. Consider the coefficients corresponding to the output sampling instants indexed by $m-1$ and $m$. Assume we keep in memory at this point the value $(e^{\alpha_kT_y})^{m-1}$ and $e^{\alpha_kT_y}$. Thus, in order to compute $(e^{\alpha_kT_y\epsilon_{y,m}})^m$, we can write the following
\begin{align}
\label{eq:AdaptiveRate_computation}
(e^{\alpha_kT_y\epsilon_{y,m}})^m&=(e^{\alpha_kT_ym})^{\epsilon_{y,m}}\nonumber \\
&=(e^{\alpha_kT_y}e^{\alpha_kT_y(m-1)})^{\epsilon_{y,m}}.
\end{align}

This calculation firstly requires the computation of $(e^{\alpha_kT_y})^m$ which can be done recursively followed by the operation of raising this value to the power of $\epsilon_{y,m}$. The same applies to the coefficients $(e^{\alpha_kT_x\epsilon_{x,n}})^n$. The added computational complexity, apart from the recursion, is then limited to performing this exponentiation. Alternatively, if the nonuniform time instants are given by $T+\varepsilon$, it is possible to compute, in addition to $(e^{\alpha_kT_y})^m$, the value $e^{\alpha_k\varepsilon m}$. This may be, in principle, less computationally efficient as $m$ grows larger. In this case, the computation could be reduced by using the principle of addition-chain exponentiation \cite[Chapter 4]{Kunuth:1998aa}. In Sections \ref{section:FirstOrder}, \ref{section:SecondOrderSystems}, and \ref{section:NOrderSystems}, we show in detail for the different cases how the summation in (\ref{eq:FOexample}) can be computed recursively.

\subsection{Input and Output Rate Ratio: Computational Complexity}
The approach presented in this paper can accommodate nonuniform input and output sampling rates. However, there are certain nonuniform input and output sampling instants that are particularly amenable to efficient computation. In order to see this, note first that the summation in (\ref{eq:NonuniformSRC_Separation}) can be computed recursively: we can use the result of the summation used to compute $y[m-1]$ for the output sample $y[m]$. For each $m$, there exists some $n_m$ such that $n_m\leq\lambda_{m,n_m}$ and $1+n_m>\lambda_{m,1+n_m}$. Thus, the additional number of terms in the summation for each output sample $m$ is given by
\begin{equation}
M_{m}:=\lambda_{m,n_{m}}-\lambda_{m-1,n_{m-1}}.
\end{equation}

\begin{figure}[thpb]
\centering
\includegraphics[width=1\columnwidth]{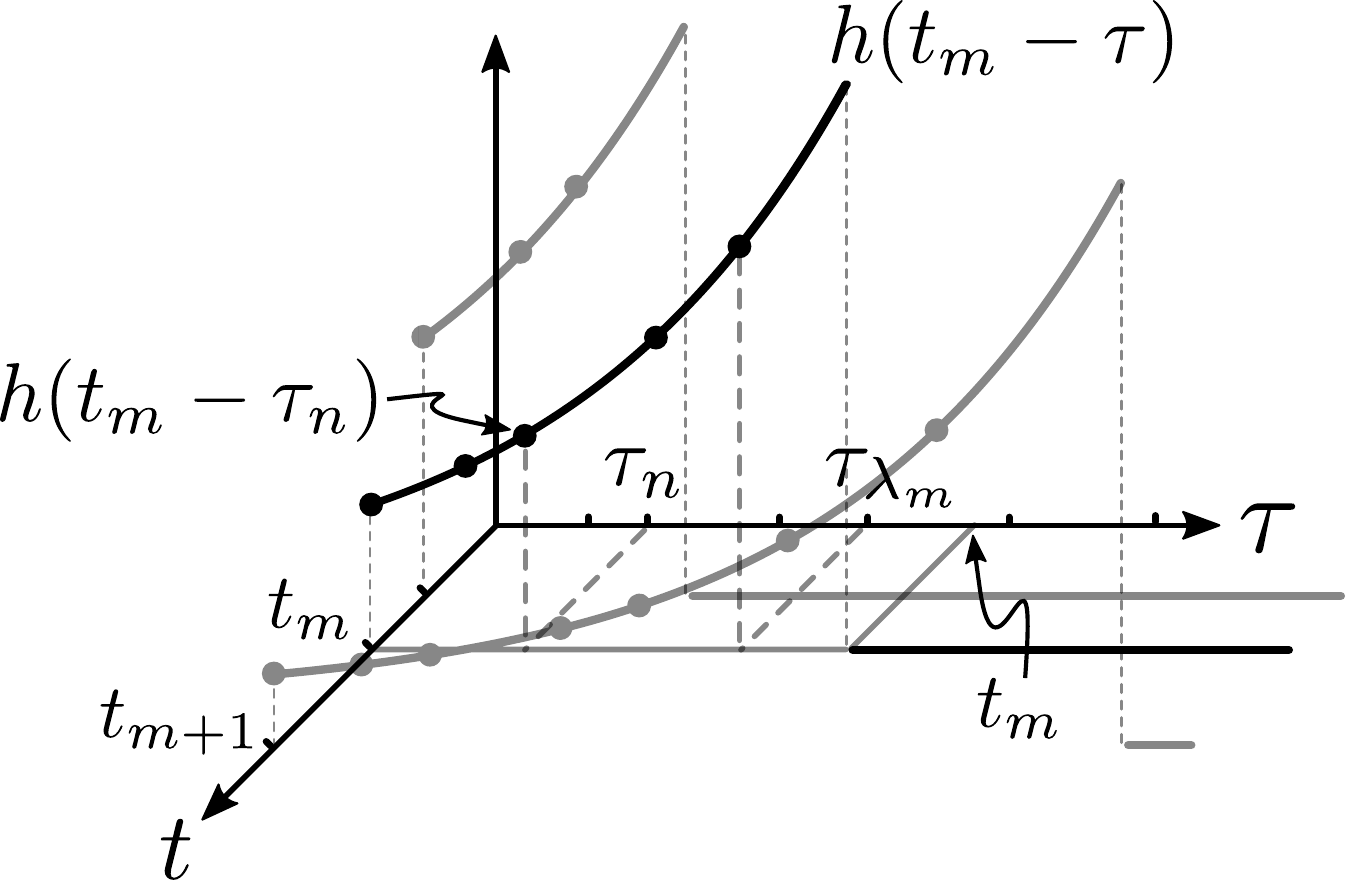}
\caption{Interpretation of a nonuniform sampling rate converter by considering samples of $h(t-\tau)$ along both time axes, i.e., $\{h(t_m-\tau_n)\}$, where $h(t)=e^{-\alpha t}u(t)$.}
\label{fig:Example}
\end{figure}

In order to illustrate this recursion, consider the function $h(t-\tau)$. We can interpret (\ref{eq:NonUniSRC}) as a linear combination of its sampled versions along both time axes (corresponding to the input or output time variables). Fig.~\ref{fig:Example} depicts this function for the system in (\ref{eq:FOexample}) with $\alpha_k<0$. It shows that, for each $m$, we have samples of the function $h(t_m-\tau)$ at instants $\{\tau_n\}_{\tau_n\leq t_m}$. However, due to the separation property, we can arrange the computation as in (\ref{eq:NonuniformSRC_Separation}). Thus, the summation performed to obtain $y[m-1]$ can be reused by adding the terms $x[n]h_2(\tau_n)$ corresponding to $t_m-1<\tau_n\leq t_{m}$. In other words, the number of terms to compute this recursion depends on how many input sampling instants fall between two consecutive output sampling instants. This number is precisely $M_m$, which can be equivalently expressed as $M_m=|\mathfrak{T}_m|$ where $\mathfrak{T}_m=\{\tau_n:t_m-1<\tau_n\leq t_{m}\}$. Notice that $M_m$ may be relatively small depending on the input and output sampling instants.

We can illustrate this more formally by taking as an example uniform input and output sampling rates. Let us first decompose the conversion ratio as follows
\begin{equation} 
\frac{T_y}{T_x}=l+r
\end{equation}
where $l$ is a nonnegative integer and $0\leq r<1$. Then, the additional number of terms in the summation reduces to
\begin{align}
\label{eq:SummationTerms}
M_{m}&=\lambda_{m}-\lambda_{m-1}\nonumber \\
&=l+\lfloor mr\rfloor-\lfloor (m-1)r\rfloor\nonumber \\
&=l+\Delta_{m}
\end{align}
where $\Delta_{m}\in\{0,1\}$. Note that if $T_y<T_x$, then $l=0$ and depending on the index $m$, $\Delta_m$ determines if there is an additional term in the summation or not. In other words, if the output rate is faster than the input rate, it is only needed to compute at most one term in the summation. Similarly, when both sampling rates are close, we may also have $l=1$. This reduces the number of terms in the summation significantly. 

A particular case involving nonuniform samples that may be specially amenable to efficient computation corresponds to uniform input rate and nonuniform output rate. In practice, this situation can arise when there exists some drift in any two interconnected digital systems with different clocks operating at uniform nominal rates. Then, the output rate can be nonuniform relative to the input rate. In other words, the conversion ratio varies with time as a result of this drift. These samples can be used for further processing as discussed here or for subsequent nonuniform reconstruction \cite{Yao:1967aa,Feichtinger:1994aa,Margolis:2008aa}.

In this scenario, the computational efficiency comes from the fact that the exponentiation is performed solely for the factor $h_1(t_m)$. We may have more coefficients to update in the summation, which are determined by $\lfloor mT_y\epsilon_{y,m}/T_x\rfloor$, that would only require multiplications without the need of exponentiations. In the next sections, we will illustrate in detail how this computation can be arranged and show how these properties apply to different systems.
 
\section{First-Order Systems: \\Single pole on the Real Axis}
\label{section:FirstOrder}
For ease of notation, we denote the nonuniform input sampling instants by $nT_x$ instead of $nT_x\epsilon_{x,n}$ and similarly for the output sampling instants. The reader can still assume that $T_x$ can vary from sample to sample as presented in previous sections. We will also adhere to this notation in Sections \ref{section:NOrderSystems} and \ref{section:SecondOrderSystems}.

Let us first consider a first-order system with the impulse response
\begin{equation}
h(t)=e^{-\alpha t}u(t)
\end{equation}
with $\alpha>0$ and Laplace transform $H(s)=1/(s+\alpha)$ for $\Re(s)>-\alpha$. Then, the time-varying discrete-time system takes the form
\begin{align}
\label{eq:FirstOrderDecayExp}
y[m]&=\sum_{n=0}^{\lambda_m}x[n]h(mT_y-nT_x)\nonumber \\
&=\sum_{n=0}^{\lambda_m}x[n]e^{-\alpha(mT_y-nT_x)}\nonumber\\
&=(e^{-\alpha T_y})^m\sum_{n=0}^{\lambda_m}x[n](e^{+\alpha T_x})^n\nonumber\\
&=c_y^m\sum_{n=0}^{\lambda_m}x[n]c_x^n=c_y^{m}g[m].
\end{align}
The constants $c_y$ and $c_x$ depend solely on the respective sampling periods $T_y$ and $T_x$. Moreover, the computation in (\ref{eq:FirstOrderDecayExp}) can be performed recursively
\begin{equation}
\label{eq:Recursive_g}
g[m+1]=g[m]+q[m+1]
\end{equation}
where the function $q[m+1]$ takes the form
\begin{equation}
\label{eq:q_LTVsystem}
q[m+1]=
	\begin{cases} 
      0 & \lambda_m<\lfloor (m+1)\frac{T_y}{T_x}\rfloor-1 \\
      \sum_{n=\lambda_m+1}^{\lfloor (m+1)\frac{T_y}{T_x}\rfloor}x[n]c_x^n & \lambda_m\geq\lfloor (m+1)\frac{T_y}{T_x}\rfloor-1.
   \end{cases}
\end{equation}

There are two important distinctions to be made. First, if $T_y<T_x$, i.e., the rate of output samples is higher than the rate of input samples, there will often be cases where $q[m+1]=0$. If $q[m+1]\neq0$, it will only consist of, at most, the term $q[m+1]=x[\lambda_m+1]c_x^{\lambda_m+1}$. Second, if the rate of output samples is slower than the input rate, then there will always be at least one term in the summation to compute $q[m+1]$ as shown in (\ref{eq:SummationTerms}). 

\subsection{Computation Ordering}
The output samples are given by $y[m]=c_y^mg[m]$ where, clearly, the factor $c_y^m$ can be recursively computed by means of one multiplication for each output sample rate and the corresponding exponentiation for nonuniform output sampling instants. The intermediate sample values $g[m]$ can also be computed in a recursive manner as $g[m]=g[m-1]+q[m]$ where $q[m]$ is given by (\ref{eq:q_LTVsystem}). Note again that the coefficients $c_x^n$ needed to generate $q[m]$ can be recursively computed and the number of coefficients needed for each output time step $M_m$ depends on the sample instants and the ratio $T_y/T_x$. Similarly, the coefficients $c_x^n$ needed to compute $q[m]$ can be obtained recursively with the corresponding exponentiations for nonuniform input sampling instants.

\begin{figure}[thpb]
\centering
\includegraphics[width=.85\columnwidth]{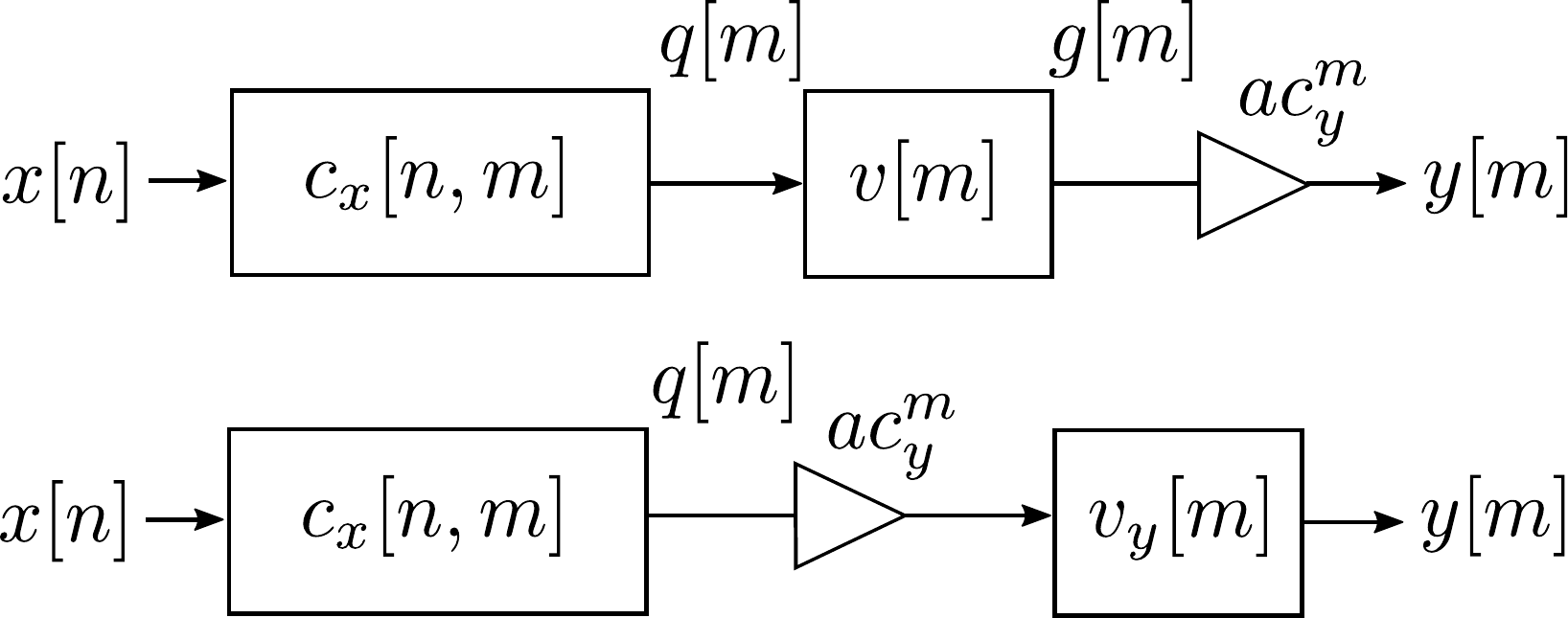}
\caption{Block-diagram representation of the time-varying filter $h[n,m]$ for an underlying first-order continuous-time system of the form $ae^{-\alpha t}u(t)$ for some $a\in\mathbb{R}$. Both systems are input-output equivalent. The interchange of the last two operations results in the bottom one having a recursive filter parametrized by $c_y$.}
\label{fig:SRC_Single_FirstOrder}
\end{figure}

We can denote the input-output relationship in (\ref{eq:Recursive_g}) by $c_x[n,m]$. Note again that if $T_y<T_x$ there will be instants at which $q[m]=0$ or, in other words $g[m]=g[m-1]$, thus no computations whatsoever are made to compute $g[m]$. If we consider uniform input and output sequences, we can compute $g[m]$ recursively using $q[m]$ as the input to a linear time-invariant system with $z$-transform $V(z)=1/(1-z^{-1})$ (see Fig.~\ref{fig:SRC_Single_FirstOrder}). The output in this case is computed by at most one addition whenever $q[m]\neq0$. Alternatively, the computation can also be rearranged in the way shown at the bottom of Fig.~\ref{fig:SRC_Single_FirstOrder}, i.e.,
\begin{equation}
\begin{split}
y[m]&=c_y^m(g[m-1]+q[m])\nonumber\\
&=c_yc_y^{m-1}g[m-1]+c_y^mq[m]\nonumber\\
&=c_yy[m-1]+c_y^mq[m].
\end{split}
\end{equation}
Thus, for an input $c_y^mq[m]$, we can express $y[m]$ as the output of a linear time-invariant system $v_y[m]$ with $z$-transform $V_y(z)=1/(1-c_yz^{-1})$. However, the ordering of operations represented by the top block diagram in Fig.~\ref{fig:SRC_Single_FirstOrder} is more convenient whenever we have instantaneous changes in the sampling rates since the parameters of $v[m]$ are independent of changes in $T_y$.

\subsection{Generalized Structure}
We can generalize the system of the previous section by considering an impulse response formed by $K$ first-order systems. These can be arranged in a parallel fashion. In particular, we have that
\begin{equation}
\label{eq:FirstOrder_IRs}
h(t)=u(t)\sum_{k=1}^{K}a_ke^{-\alpha_kt}
\end{equation}
for $\alpha_k>0$ and $a_k\in\mathbb{R}$.

\begin{figure}[thpb]
\centering
\includegraphics[width=.95\columnwidth]{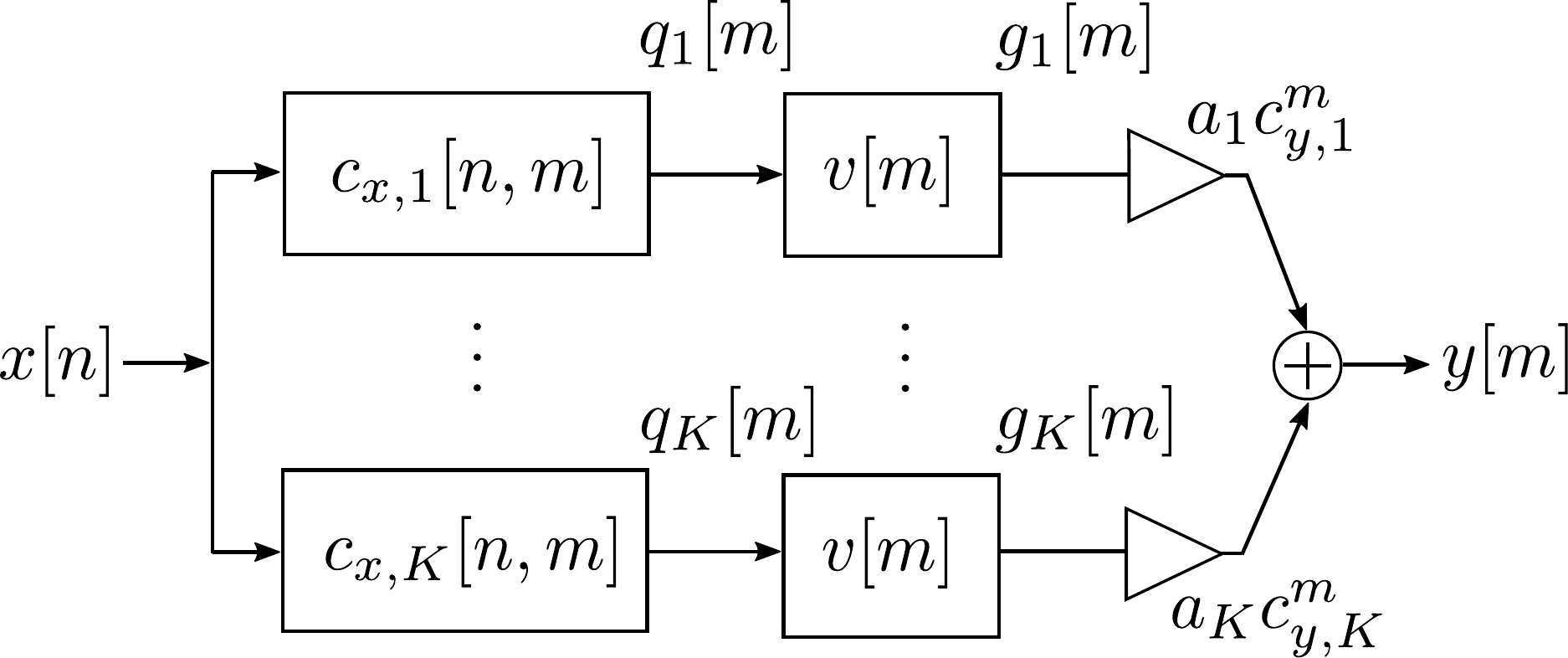}
\caption{Block-diagram representation of a nonuniform sample rate converter where the underlying continuous-time filter consists of a linear combination of first-order systems.}
\label{fig:StructureSRC}
\end{figure}

In this case, the output $y[m]$ is given by
\begin{align}
y[m]&=\sum_{n=0}^{\lambda_m}x[n]\sum_{k=1}^Ka_ke^{-\alpha_kmT_y}e^{+\alpha_knT_x}\nonumber\\
&=\sum_{k=1}^Ka_k(e^{-\alpha_kT_y})^m\sum_{n=0}^{\lambda_m}x[n](e^{+\alpha_kT_x})^n\nonumber\\
&=\sum_{k=1}^Ka_kc_{y,k}^m\sum_{n=0}^{\lambda_m}x[n]c_{x,k}^n=\sum_{k=1}^Ka_kc_{y,k}^mg_k[m].
\end{align}
The same computational principles shown before extend to this case. Again, if the input and output sequences are nonuniform, we would have to take into account the corresponding exponentiation required to update the respective coefficients. Fig.~\ref{fig:StructureSRC} shows the parallel structure of a nonuniform sampling rate converter when the underlying continuous-time filter is given by (\ref{eq:FirstOrder_IRs}).

\subsection{Computational Complexity}
In order to provide an idea of the computational demands of this algorithm, we can consider the number of additions and multiplications, the complexity of the exponentiation step, and the memory requirements. We will focus on the operations needed to generate just one output sample.

Considering a single first-order system and uniform sequences, we first have to keep in memory $a$, $c_y$, $c_y^{m-1}$, $c_x$, $c_x^{\lambda_{m-1}}$, and $g[m-1]$. In order to compute $g[m]$ we need $\mathbf{1}_{M_m\geq1}(M_m)$ additions---since there are $M_m=l+\Delta_m$ terms in the residual summation and $v[m]$ only requires one addition---and $2M_m=2(l+\Delta_m)$ multiplications. Then, we need one multiplication to update the factor $c_y^{m}$ and two multiplications to obtain $y[m]=ac_y^{m}g[m]$. Note that we have separated the factor $a$ from $c_y$ since, in the case of nonuniform output rates, this leads to a more efficient exponentiation as shown in (\ref{eq:AdaptiveRate_computation}). This gives us a total of $2(M_m+1)+1$ multiplications per output sample. 

If the output rate is uniform, the corresponding constant $a$ does not add any operation to obtain $ac_{y}^{m}$ since we can assume the values $ac_{y}^{m-1}$ and $c_{y}$ are kept in memory. 

If we have nonuniform input or output sequences, the added complexity relies on performing the corresponding exponentiations whenever the sampling instants do not fall into a uniform grid. In particular, let us denote the number of exponentiations corresponding to the coefficients $\{c_x^n\}$ as 
\begin{equation}
\label{eq:ExponentiationsX}
E_m:=\mkern-18mu\sum_{n=\lambda_{m-1}+1}^{\lambda_m}\mkern-18mu\mathbf{1}_{\epsilon_{x,n}\neq1}(\epsilon_{x,n})
\end{equation}
which, obviously, satisfies $E_m\leq M_m$. We may also have to perform one exponentiation for $(e^{\alpha_kT_y\epsilon_{y,m}})^m$ as shown in (\ref{eq:AdaptiveRate_computation}), and $E_m$ exponentiations for coefficients of the form $(e^{\alpha_kT_x\epsilon_{x,n}})^n$. This amounts to $E_m+\mathbf{1}_{\epsilon_{y,m}\neq1}(\epsilon_{y,m})$ real exponentiations per output sample.

\section{Second-Order Systems: \\Poles in Complex Conjugate Pairs}
\label{section:SecondOrderSystems}
The same principle of separation of variables is satisfied by an impulse response that takes the following form
\begin{equation}
\label{eq:IRsecondorder}
h(t)=e^{-\alpha t}e^{j\omega_o t}u(t)
\end{equation}
where $\alpha>0$ and $\omega_o\in\mathbb{R}$. The operations can be rearranged as in (\ref{eq:FirstOrderDecayExp})
\begin{align}
\label{eq:SecondOrderDecayExp}
y[m]&=\sum_{n=0}^{\lambda_m}x[n]e^{-\alpha(mT_y-nT_x)}e^{j\omega_o(mT_y-nT_x)}\nonumber\\
&=(e^{-\alpha T_y+j\omega_o T_y})^m\sum_{n=0}^{\lambda_m}x[n](e^{+\alpha T_x-j\omega_o T_x})^n\nonumber\\
&=\hat{c}_y^m\sum_{n=0}^{\lambda_m}x[n]\hat{c}_x^n=\hat{c}_y^{m}g[m]
\end{align}
where the constants $\hat{c}_y$ and $\hat{c}_x$ can be precomputed or updated if there are changes in the sampling rates. Again, this is similar to the description of first-order systems.

\begin{figure}[thpb]
\centering
\includegraphics[width=.95\columnwidth]{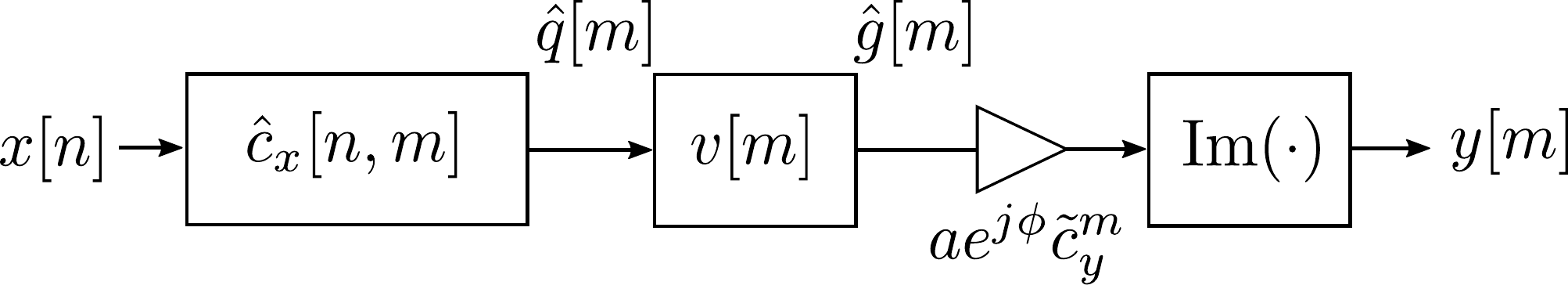}
\caption{Block-diagram representation of a nonuniform sampling rate converter where the underlying continuous-time second-order system takes the form in (\ref{eq:RealSecondOrder}). The last block takes the imaginary part of the complex input sample.}
\label{fig:SRC_Single_SecondOrder}
\end{figure}

We are interested in the case of real second-order systems that are causal and stable. These systems have an impulse response that can be expressed as
\begin{equation}
\label{eq:RealSecondOrder}
h(t)=ae^{-\alpha t}\sin(\omega t+\phi)u(t)
\end{equation}
for $a,\ \omega$ and $\phi\in\mathbb{R}$ and $\alpha>0$. In this case, the associated computation to obtain the output samples $y[m]$ can be carried out by rearranging the operations in a manner similar to the preceding, namely
\setlength{\arraycolsep}{0.0em}
\begin{eqnarray}
y[m]&{}={}&\frac{a}{2j}(e^{-\alpha T_y})^m\Big[e^{+j\phi}(e^{+j\omega T_y})^m\sum_{n=0}^{\lambda_m}x[n](e^{+\alpha T_x-j\omega T_x})^n\nonumber\\
&&{-}\:e^{-j\phi}(e^{-j\omega T_y})^m\sum_{n=0}^{\lambda_m}x[n](e^{+\alpha T_x+j\omega T_x})^n\Big]\nonumber\\
&{}={}&\frac{a}{2j}c_y^m\Big[e^{+j\phi}\tilde{c}_y^m\sum_{n=0}^{\lambda_m}x[n]\hat{c}_x^n-e^{-j\phi}(\hat{c}_y^*)^m\sum_{n=0}^{\lambda_m}x[n](\hat{c}_x^*)^n\Big]\nonumber\\
&{}={}&\frac{a}{2j}c_y^m\Big[e^{+j\phi}\tilde{c}_y^m\hat{g}[m]-(e^{+j\phi}\hat{c}_y^m\hat{g}[m])^*\Big]\nonumber\\
&{}={}&\Im(ae^{+j\phi}\hat{c}_y^m\hat{g}[m])
\end{eqnarray}
where $\tilde{c}_y=e^{-j\omega T_y}$, $\hat{c}_x=e^{(\alpha T_x-j\omega T_x)}$, and $\hat{c}_y=c_y\tilde{c}_y$. Fig.~\ref{fig:SRC_Single_SecondOrder} shows a block-diagram representation of a nonuniform sampling rate converter when $h(t)$ is a real second-order system. The operations involved are very similar to the first-order case (see Fig.~\ref{fig:SRC_Single_FirstOrder}) with the added number of real multiplications---now the coefficients are complex numbers---and the computationally inexpensive operation of keeping solely the imaginary part to generate the appropriate output sample.

If we have a sum of second-order systems, such as $h(t)=u(t)\sum_{k=1}^Ka_ke^{-\alpha_kt}\sin(\omega_kt+\phi_k)$, the generalized structure is constructed similarly to the case of first-order systems.

\subsection{Computational Complexity}
We have to keep in memory the complex values $ae^{+j\phi}$, $\hat{c}_y$, $\hat{c}_y^{m-1}$, $\hat{c}_x$, $\hat{c}_x^{\lambda_{m-1}}$, and $\hat{g}[m-1]$. Evidently, we assume that the input $x[n]$ is a real signal. The factor $\hat{c}_y^{m}$ can be computed recursively with four real multiplications and two additions. Similarly to the situation of first-order systems, we factorize $ae^{+j\phi}$ as in (\ref{eq:AdaptiveRate_computation}) to perform the exponentiation more efficiently. The term $\hat{g}[m]$ can also be computed recursively by using $\hat{g}[m-1]$. This requires the calculation of $M_m$ coefficients of the form $\hat{c}_x^n$ that can also be computed recursively with four multiplications and two additions per coefficient. This amounts to $6M_m=6(l+\Delta_m)$ real multiplications and $2M_m+2\mathbf{1}_{M_m\geq1}(M_m)$ additions in order to obtain $\hat{g}[m]$. As a result, we have a total of $6(M_m+1)+4$ real multiplications---since we are only interested in the imaginary part---and $2M_m+2\mathbf{1}_{M_m\geq1}(M_m)+3$ additions per output sample.

If the input or output rate is always uniform, we could combine the constant factor $ae^{j\phi}$ into $\hat{c}_y$ or $\hat{c}_x$---depending on what rate is uniform. This results in a slight reduction of memory and computational requirements. If we have nonuniform input and output sequences, updating the coefficients requires the corresponding exponentiations whenever the sampling instants do not fall into a uniform grid. Thus, using the notation in (\ref{eq:ExponentiationsX}), this would require $E_m+\mathbf{1}_{\epsilon_{y,m}\neq1}(\epsilon_{y,m})$ complex exponentiations.

\section{Repeated Poles on the Real Axis}
\label{section:NOrderSystems}
We can also consider a system whose Laplace transform consists of repeated poles on the real axis. Transfer functions with repeated poles---real or in complex conjugate pairs---rarely appear in practice. However, we include the development here for illustrative purposes and to show how the separation property can be used to arrange the computation in other systems. In principle, a similar approach can be applied to the case of repeated pairs of complex conjugate pairs. The impulse response in the case of real poles with multiplicity can then be expressed as
\begin{equation}
\label{eq:IRmultiplicity}
h(t)=at^Ne^{-\alpha t}u(t)
\end{equation}
for $\alpha>0$, $N\geq0$, and $a\in\mathbb{R}$.

\begin{figure}[thpb]
\centering
\includegraphics[width=.95\columnwidth]{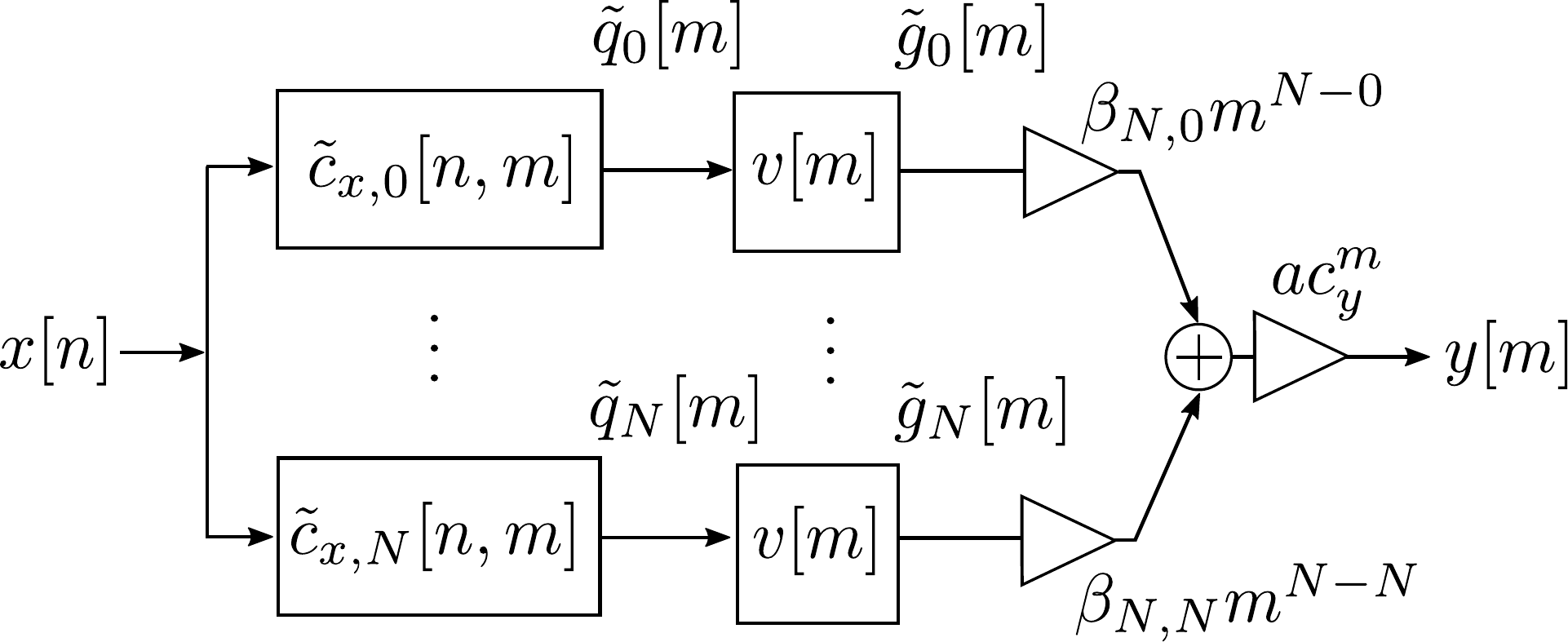}
\caption{Block-diagram representation of a nonuniform sampling rate converter where the underlying continuous-time system consists of real poles with multiplicity as in (\ref{eq:IRmultiplicity}).}
\label{fig:N_OrderSystem}
\end{figure}

We can then write the output of the time-varying discrete-time sampling rate converter as 
\begin{equation}
\begin{split}
y[m]&=a\sum_{n=0}^{\lambda_m}x[n](mT_y-nT_x)^N(e^{-\alpha T_y})^m(e^{+\alpha T_x})^n\nonumber\\
&=ac_y^m\sum_{n=0}^{\lambda_m}x[n]c_x^n\sum_{k=0}^{N}\binom{N}{k}(mT_y)^{N-k}(-nT_x)^{k}\nonumber \\
&=ac_y^m\sum_{k=0}^{N}\binom{N}{k}T_y^{N-k}m^{N-k}\sum_{n=0}^{\lambda_m}x[n]c_x^n(-nT_x)^{k}.
\end{split}
\end{equation}
If the input and output sequences are uniform, the computation can be arranged as 
\begin{equation}
y[m]=ac_y^m\sum_{k=0}^{N}\binom{N}{k}T_y^{N-k}T_x^km^{N-k}\sum_{n=0}^{\lambda_m}x[n]c_x^n(-n)^{k}.
\end{equation}
By denoting the values $\beta_{N,k}=\binom{N}{k}T_y^{N-k}(-T_x)^k$, which can be precomputed and stored in memory, we can write the output as
\begin{equation}
y[m]=ac_y^m\sum_{k=0}^{N}\beta_{N,k}m^{N-k}\tilde{g}_k[m]
\end{equation}
where $\tilde{g}_k[m]$ can be recursively computed as $\tilde{g}_k[m+1]=\tilde{g}_k[m]+\tilde{q}_k[m+1]$. The definition of $\tilde{q}_k[m]$ is analogous to (\ref{eq:q_LTVsystem}) with the added factor $n^k$. As shown before, the number of terms involve in the summation involved in $\tilde{q}_k[m]$ depends on the ratio $T_y/T_x$. This rearrangement of operations gives us the structure shown in Fig.~\ref{fig:N_OrderSystem}.

Notice that if the multiplicity corresponds to $N=0$, the system in Fig.~\ref{fig:N_OrderSystem} reduces to the one shown in Fig.~\ref{fig:SRC_Single_FirstOrder} and, obviously, all the considerations developed in that case applied equally to this case.

If the input and output sequences are nonuniform, we need to introduce different definitions. In particular, we need to substitute $\beta_{N,k}$ by
\begin{equation}
\tilde{\beta}_{N,k,m}=\binom{N}{k}(T_y\epsilon_{y,m})^{N-k}
\end{equation}
which explicitly depends on the corresponding output sampling instant. Consequently, the term $\tilde{g}_k[m]$ takes the form
\begin{equation}
\tilde{\mathfrak{g}}_{k}[m]=\sum_{n=0}^{\lambda_m}x_nc_x^n(-nT_x\epsilon_{x,n})^{k}.
\end{equation}

\subsection{Generalized Structure}
It is possible to generalize the previous case in a straightforward way to systems with an impulse response of the form
\begin{equation}
\label{eq:IRmultiplicity_parallel}
h(t)=\sum_{i=1}^La_it^{N_i}e^{-\alpha_i t}u(t).
\end{equation}
In this case, we can write
\begin{equation}
y[m]=\sum_{i=1}^{L}a_ic_{y,i}^m\sum_{k=0}^{N_i}\beta_{N,k}^{i}m^{N_i-k}\tilde{g}_k^i[m].
\end{equation}

\begin{figure}[thpb]
\centering
\includegraphics[width=.95\columnwidth]{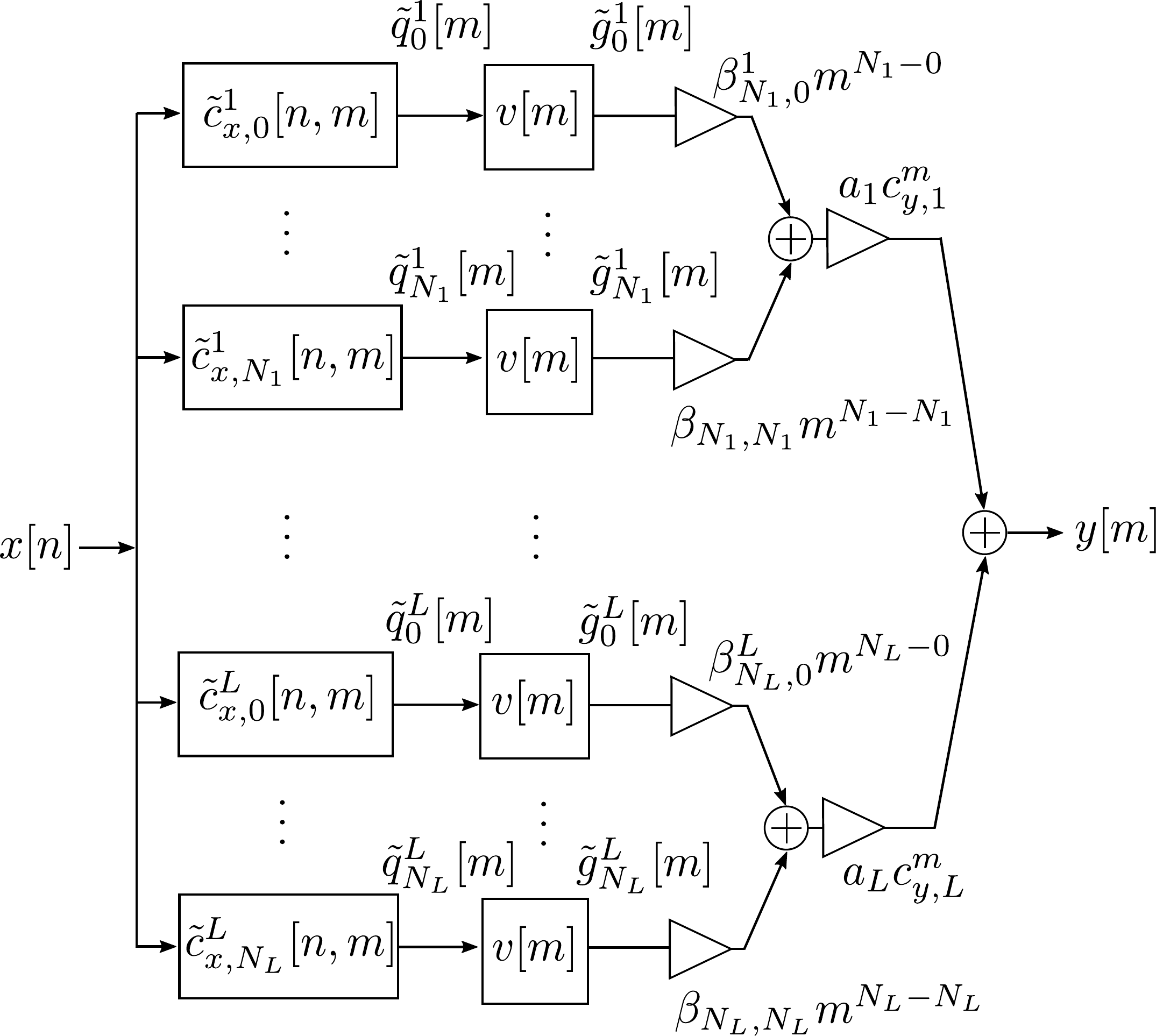}
\caption{Block-diagram representation of a nonuniform sampling rate converter where the underlying continuous-time systems all consist of real poles with multiplicity as in (\ref{eq:IRmultiplicity_parallel}).}
\label{fig:N_OrderSystemS}
\end{figure}

The resulting structure can be readily combined in the manner shown in Fig.~\ref{fig:N_OrderSystemS}.

\subsection{Computational Complexity}
In this case, irrespective of having uniform or nonuniform sequences, it is required at some stage to compute powers of the form $a^\nu$ for some $a\in\mathbb{R}$ and a positive integer $\nu$. The naive approach would require $\nu-1$ multiplications. However, this can be further reduced by resorting to addition-chain exponentiation \cite[Chapter 4]{Kunuth:1998aa}. In this section, we take the worst-case scenario by considering this naive approach.

Consider the system with impulse response shown in (\ref{eq:IRmultiplicity}). In order to generate one output sample $y[m]$, we first need to store the $(N+1)$ coefficients $\{\beta_{N,k}\}_{k=0}^N$ as well as $a$, $c_y$, $c_y^{m-1}$, $c_x$, and $c_x^{\lambda_{m-1}}$. Likewise first- and second-order systems, keeping in memory these last four factors makes the computation more efficient for nonuniform sequences.

In order to compute $\tilde{g}_k[m]$, we need $\mathbf{1}_{M_m\geq2}\cdot(M_m-1)+1$ additions and $2M_m$ multiplications. For each output sample, we also have to compute $\{m^k\}_{k=0}^N$ and $\{n^k\}_{k=0}^N$ for $n=1+\lambda_{m-1},\ldots,\lambda_m$ which requires $(M_m+1)N(N-1)/2$ multiplications. Before combining all the channels (see Fig.~\ref{fig:N_OrderSystem}), we need to multiply by the corresponding factors $\beta_{N,k}m^{N-k}$. This results in $2N+1$ multiplications. The output of each channel is added together by means of $N$ additions to generate an output sample. Thus, the computations per output sample come down to $\mathbf{1}_{M_m\geq2}\cdot(M_m-1)+1+N$ additions and $(M_m+1)N(N-1)/2+2M_m+2N+1$ multiplications. Note that raising a number to zero or one, or multiplying by unity are not considered multiplications. The added complexity regarding nonuniform input or output sequences lies in updating the coefficients $\tilde{\beta}_{N,k,m}$, which requires $N(N-1)/2+(N-1)$ multiplications, and the corresponding exponentiations to update $c_x^n$ and $c_y^m$

If our system can be expressed as the sum of $L>0$ distinct systems of the form shown in (\ref{eq:IRmultiplicity}), the computational complexity is simply increased by a factor $L$.

\section{Example}
\label{section:example}
We consider a third-order lowpass Butterworth filter split into a parallel structure consisting of a first- and a second-order section. In this scenario, we assume that errors in the acquisition of the digital signal, with a nominal rate of 48 kHz, have led to nonuniform periodic sampling \cite{Nikaeen:2009aa}. The output rate is taken to be 44.1 kHz. In particular, we can assume, for example, that the input instants are given by $\tau_n:=nT_x+\delta_nT_x$ where
\begin{equation}
\delta_n:=
\begin{cases}
1/4, \textrm{ $n$ even}\\
1/5,\textrm{ $n$ odd}
\end{cases}
\end{equation}
for $T_x = (1/48)\cdot10^{-3}$~s and $n\in\mathbb{Z}$. The output sampling instants are simply $t_m := mT_y$ for $T_y = (1/44.1)\cdot10^{-3}$~s and $m\in\mathbb{Z}$. Following our notation, we can also write $\epsilon_{x,n} = 1+\delta_n/n$ for $n\neq0$. 

The cutoff frequency of the lowpass Butterworth filter is set to 20 kHz. The transfer function then takes the form
\begin{equation}
\label{eq:butter_transfer}
H(s) = \frac{A}{(s-s_o)(s-s_1)(s-s^*_1)}
\end{equation}
where the gain and poles and given in Table \ref{table:filter} and the region of convergence is for $\Re(s)>\Re(s_1)$.

The parallel structure can be derived from a partial fraction expansion of (\ref{eq:butter_transfer}). For the sake of illustration, let us focus on the first-order section. In this case, the impulse response is given by
\begin{equation}
\label{eq:butter_impulse}
h_0(t):=ae^{s_0 t}u(t)
\end{equation}
where $a$ can also be found in Table \ref{table:filter}.
The output of this section---assuming that the input signal is zero for negative time indexes---is
\begin{align}
y^{(0)}[m]=ac_y^m\sum_{n=0}^{\lambda_m}x[n]c_x^n
\end{align}
where $c_y^m=(e^{-\alpha T_y})^m$ and 
\begin{equation}
c_x^n =
\begin{cases}
e^{+\alpha (T_x/4)}(e^{+\alpha T_x})^n,\textrm{ $n$ even}\\
e^{+\alpha (T_x/5)}(e^{+\alpha T_x})^n,\textrm{ $n$ odd}.
\end{cases}
\end{equation}
for $n>0$. Due to the structure of the input and output sampling instants, both $c_y^{m}$ and $c_x^{n}$ can be computed recursively without repeated exponentiations. In terms of memory, it just amounts to storing 6 real values including $a$.

In this case, the output rate is slower than the input rate, thus $q[m]$ always requires computations of the summation terms. In particular, the number of terms in each summation is $\lambda_{m,n}=\lfloor m48/44.1-\delta_n\rfloor$ for $m\geq1$. It is straightforward to see that $\lambda_m = \lfloor m48/44.1-1/4\rfloor$ for $m\geq1$. Moreover, it can be easily shown---from the properties of the floor function---that the number of additional terms for each output sample is bounded, i.e., $1\leq M_{m+1}\leq2$ for $m\geq1$. Thus, the number of additions and multiplications is also bounded per output sample. The computation of $q[m]$ requires either 3 or 6 multiplications for each $m$. We also need to store the previous summation result and add it to $q[m]$: this is the recursion expressed as $v[m]$ resulting in $g[m]$. Lastly, we have 2 more multiplications due to $ac_y^{m}g[m]$. Thus, for each output sample, the first-order system amounts to either 6 or 9 multiplications, either 1 or 2 additions, and 6 values previously stored in memory.

\begin{table}[!t]
\renewcommand{\arraystretch}{1.3}
%if using array.sty, it might be a good idea to tweak the value of
%extrarowheight as needed to properly center the text within the cells
\caption{Parameters of the transfer function and the first-order section of the impulse response of the Butterworth filter specified in (\ref{eq:butter_transfer}) and (\ref{eq:butter_impulse}), respectively.∂}
\label{table:filter}
\centering
% Some packages, such as MDW tools, offer better commands for making tables
% than the plain LaTeX2e tabular which is used here.
\begin{tabular}{||c|c||}
\hline
$A$ & 1984401707539188.5\\
\hline
$s_0$ & $-125663.70614359$\\
\hline
$s_1$ & $-62831.8530718 + 108827.96185405j$\\
\hline\hline
$a$ & $125663.70614360292$\\
\hline
\end{tabular}
\end{table}

\begin{figure}[t]
\centering
\includegraphics[width=.8\columnwidth]{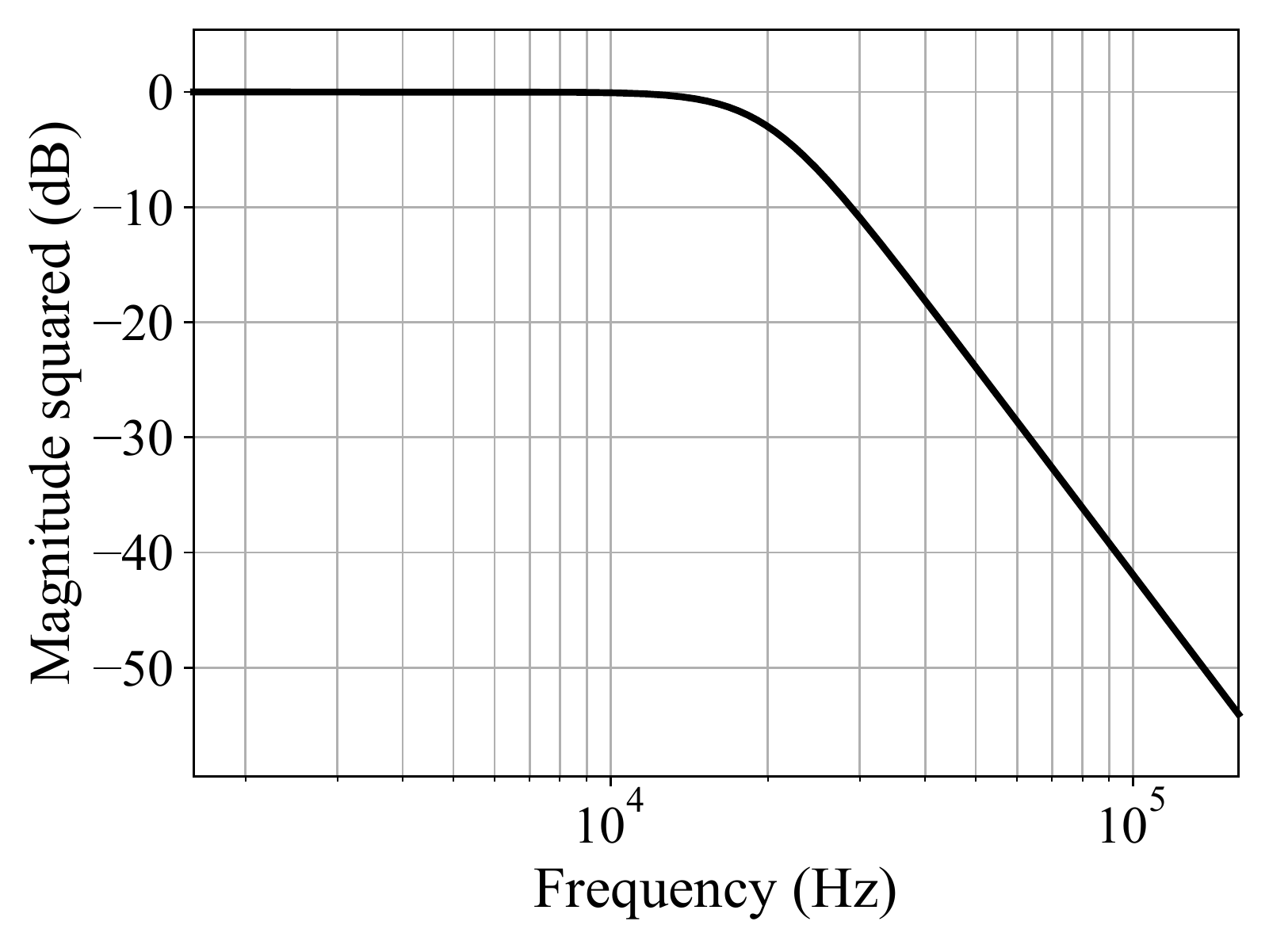}
\caption{Magnitude squared of the frequency response of the lowpass Butterworth filter in (\ref{eq:butter_transfer}).}
\label{fig:filter}
\end{figure}

\section{Conclusion}
\label{section:conclusion}
We have presented a discrete-time sampling rate converter that accommodates input and output sequences at nonuniform rates. This approach exploits the separation property of the underlying continuous-time filter. In particular, we focused our analysis on proper rational transfer functions with distinct poles since they can be readily decomposed into first- and second-order systems operating in parallel. We also showed how our approach can be applied to transfer functions with repeated real poles. 

Our approach can be particularly useful in a variety of contexts, e.g., sampling rate conversion between ideally uniform sampling rates that are subject to drift or mismatch between different clock domains. The added benefit is that the proposed systems can adapt efficiently to rapid variations in these sampling rates. This adaptive characteristic relies on performing exponentiations which could be made more efficient by exploiting the corresponding algorithmic methods of computation or even using dedicated hardware resources.

\ifCLASSOPTIONcaptionsoff
  \newpage
\fi

\end{document}